\documentclass[aps,prd,noshowpacs,preprintnumbers,nofootinbib,floatfix,onecolumn]{revtex4}

\usepackage{bm}
\usepackage{latexsym}
\usepackage{dcolumn}
\usepackage{amsfonts,amssymb,amsmath}
\usepackage{graphicx,epsfig}
\usepackage{psfrag}

\def\eq#1{{Eq.~(\ref{#1})}}

\def\mes#1#2{\int_{#2} d^{#1}x\, \sqrt{-g}\, }

\begin{document}


\title{HOLOGRAPHY OF  GRAVITATIONAL ACTION FUNCTIONALS}
\author{A.Mukhopadhyay}
\email{ayan@mri.ernet.in}
\affiliation{Harish-Chandra Research Institute, Chhatnag Road,
Jhunsi, Allahabad 211 019, India.\\}

\author{T. Padmanabhan}
\email{paddy@iucaa.ernet.in}
\affiliation{IUCAA, Post Bag 4, \\Ganeshkhind, Pune  411 007, India}

\date{\today}


\begin{abstract}
Einstein-Hilbert (EH) action can be separated into a bulk and a surface term, with a specific (``holographic") relationship between the two, so that either can be used to extract information about the other.  The surface term  can also  be interpreted as the entropy of the horizon in a wide class of spacetimes. Since EH action is likely to just the first term in the derivative expansion of an effective theory, it is interesting to ask whether these features continue to hold for more general gravitational actions. We provide a comprehensive analysis of  lagrangians of the form 
$\sqrt{-g}L=\sqrt{-g}Q_a^{\phantom{a}bcd}R^a_{\phantom{a}bcd}$, in which $Q_a^{\phantom{a}bcd}$ is a tensor with the symmetries of the curvature tensor, made from metric and curvature tensor and satisfies the condition $\nabla_cQ_a^{\phantom{a}bcd}=0$, and show that they share these features. The Lanczos-Lovelock lagrangians are a subset of these in which $Q_a^{\phantom{a}bcd}$ is a homogeneous function of the curvature tensor.  They are all holographic, in a specific sense of the term, and --- in all these cases --- the surface term can be interpreted as the horizon entropy. The thermodynamics route to gravity, in which the field equations are interpreted as $TdS=dE+pdV$, seems to have greater degree of validity than the field equations of Einstein gravity itself.
The results suggest that the holographic feature of EH action could also serve as a new symmetry principle in constraining the semiclassical corrections to Einstein gravity.  The implications are discussed.

\end{abstract}

\maketitle

\section{Introduction}\label{sec:intro}

Holography, in different guises, has been an attractive and influential concept in high energy physics \cite{holoorig}. This term is used in different contexts to indicate different features, with the common thread being the existence of a relation between physics in $(D-1)$ dimensional space (``surface", $\partial \mathcal{V}$) and physics in the $D$ dimensional space (``bulk", $\mathcal{V}$). In this paper, we will deal with a class of action functionals describing gravity which have a surface  contribution from  $\partial \mathcal{V}$ and bulk contribution from $\mathcal{V}$ with very specific relationship between the two. 
In particular, the form of one determines the other and the classical dynamics can be obtained from either with suitably defined variational principles. Since this is quite in the spirit of holography, we shall use this term to describe such actions; that is, \textit{we will call an action functional holographic if it can be decomposed to a surface and bulk term in a natural manner, with a specific inter-relationship that allows us to determine one from the other}. It is known that \cite{ehholography} Einstein-Hilbert action as well as some of its generalisations are indeed holographic. We will provide a  comprehensive analysis of these results
and show that all Lanczos-Lovelock lagrangians \cite{love} exhibit  holography.

In the case of Einstein-Hilbert action with the lagrangian $L_{EH}[\partial^2g, \partial g, g]$, the separation into bulk and surface terms
$L_{EH}\sqrt{-g}=L_{bulk}[\partial g,g]+L_{sur}[\partial^2g, \partial g, g]$ is quite obvious because there exists a $L_{bulk}$ (the usual $\Gamma^2$ lagrangian) which is \textit{independent} of second derivatives of the metric. But when we consider more general lagrangians, involving higher powers of curvature, say, it is not possible to effect such a simple separation. In fact, no $L_{bulk}$ which is 
independent of second derivatives of the metric, will exist for such lagrangians. What is remarkable, however, is that there is  indeed a natural way of extending the results 
\cite{tpconf} obtained for Einstein-Hilbert lagrangian to all lagrangians of the form 
$\sqrt{-g}L=\sqrt{-g}Q_a^{\phantom{a}bcd}R^a_{\phantom{a}bcd}$ in which $Q_a^{\phantom{a}bcd}$ is a tensor with the symmetries of the curvature tensor, made from metric and curvature tensor and satisfies the condition $\nabla_cQ_a^{\phantom{a}bcd}=0$. The Lanczos-Lovelock lagrangians are a subset of these in which $Q_a^{\phantom{a}bcd}$ is a homogeneous function of the curvature tensor. They are all holographic, in the sense of the term defined above, and will be the focus of our attention in this paper. 

The motivation for this analysis is three-fold: 
First, Lanczos-Lovelock lagrangian has an interesting geometrical structure and has been extensively investigated in the literature \cite{lovelockprev}. It would be nice to add to this study new features and new perspectives on previous results. As it sometimes happens, the study of a general structure sheds light on the peculiar features of a special case and here, the study of holographic properties of Lanczos-Lovelock lagrangians helps to understand the holography of Einstein-Hilbert action.  

Second, there is a point of view, shared by many, that Einstein-Hilbert action is just the first term in the derivative expansion in a low energy effective theory. At present, we have no general prescription which allows us to restrict the form of higher order quantum corrections to gravity. The only known low energy symmetry (under diffeomorphism) allows for a wide choice of correction terms. There is some evidence from string theory  that not all these choices are realised in nature. Any extra symmetry of Einstein-Hilbert action (like the action being holographic) will allow us to restrict higher order correction terms and is worth investigating. 

Third, and probably most interesting, motivation comes from the relation between gravity and thermodynamics. The surface term in Einstein-Hilbert action has a clear thermodynamic interpretation and will lead to the entropy of the horizon in a wide class of spacetimes (for a recent review, see ref. \cite{tpreview}). The notion of the horizon entropy can be generalised to an arbitrary, generally covariant lagrangian \cite{noether}. We will show that the surface term which we get, in the class of lagrangians of the form 
$\sqrt{-g}L=\sqrt{-g}Q_a^{\phantom{a}bcd}R^a_{\phantom{a}bcd}$,  leads directly to the same  entropy. In addition to being a interesting result by itself, this allows one to interpret the equations describing the gravity, including the higher derivative corrections, in thermodynamic terms \cite{tpconf}. This approach is in the spirit of what could be called the Sakharov paradigm \cite{sakharov}, in which Einstein's equations are considered similar to those describing the equations of elasticity in solid state physics. (For some of the previous attempts in the same spirit, see ref. \cite{many}). It was known that \cite{ss1}, in the case of spherically symmetric spacetimes with horizons, standard Einstein's equations can be explicitly expressed as $TdS=dE+pdV$; recently, this result has been extended to all  Lanczos-Lovelock lagrangians \cite{ss2}. This is remarkable because it makes the thermodynamic paradigm as more fundamental than a specific set of field equations. Einstein's equations will be modified by quantum corrections but some thermodynamic relation like $TdS=dE+pdV$ might remain valid to all orders \cite{tpconf,newper}.  From a practical point of view this may not seem dramatic since $S,E$ etc will pick up quantum corrections but it certainly has deep conceptual significance. 

The plan of the paper is as follows: Since we will be dealing with actions which involve second and higher derivatives of dynamical variables, we shall briefly discuss some issues related to such actions, in a simple setting, in Sec. \ref{sec:toy}. Then we will proceed to derive the holographic relationship in a wide class of actions [Sec. \ref{sec:actholo}] after briefly reviewing the Einstein-Hilbert case. In Sec \ref{sec:entro} we will show that the surface term obtained in our approach correctly gives the entropy of the horizons, thereby strengthening the thermodynamic interpretation. Sec. \ref{sec:conclu} summarises the conclusions. 

\section{Warm-up: Toy model with higher derivative action}
\label{sec:toy}

Einstein-Hilbert action and  Lanczos-Lovelock actions which we will discuss in the paper contain second derivatives of dynamical variables but their equations of motion do not have higher order terms.
The purpose of this section is to demystify this  aspect in a simple context and show
how one can a construct large family of lagrangians involving second derivatives of dynamical variables but with the resulting equations still remaining th second order in time.

Consider  a dynamical variable  $q(t)$ in point mechanics described by a lagrangian $L_q(q,\dot q)$.
 Varying the action obtained from integrating this Lagrangian
in the interval $(t_1,t_2)$ and keeping $q$ fixed at the endpoints, gives the Euler Lagrange equations for the system $(\partial L_q/\partial q)=dp/dt$, where 
we have defined
a function $p(q,\dot q)\equiv (\partial L_q/\partial \dot q)$. (The subscript $q$ on $L_q$ is an indicator of the variable that is kept fixed at the end points.) The lagrangian contains only up to first derivatives of the dynamical variable and the equations of motion are --- in general --- second degree in the time derivative. 

When the lagrangian $L_q$  depends on $\ddot q$ as well, the theoretical formulation becomes more complicated. For example, if the equations of motion become higher order, then more initial conditions are required to pose a well-defined initial value problem and the corresponding definition of path integral in quantum theory, using the lagrangian, is nontrivial \cite{ps1}. 
Interestingly enough, there exists a wide class of lagrangians $L(\ddot q,\dot q,q)$ which depend on $\ddot q$ but still lead to equations of motion which are only second order in time. We will now  motivate and analyse this class which will lead to the holographic actions in field theory.

To do this, let consider the following question: We want to modify the lagrangian $L_q$ such that  the same equations of motion are obtained when  --- instead of fixing $q$ at the end points --- we  fix some other (given) function
 $C(q,\dot q)$  at the end points. This is easily achieved by modifying the lagrangian by adding a term
$-df(q,\dot q)/dt$ which depends on $\dot q$ as well. (The minus sign is just for future convenience.) 
The new lagrangian is:
\begin{equation}
L_C(q,\dot q,\ddot q)=L_q(q,\dot q)-\frac{df(q,\dot q)}{dt}
\end{equation}
We want this lagrangian $L_C$ to lead to the same equations of motion as $L_q$, when some given function
 $C(q,\dot q)$  is held fixed at the end points. We assume $L_q$ and $C$ are given and we need to find $f$. The standard variation gives
 \begin{equation}
\delta A_C =\int_{t_1}^{t_2} dt\left[\left(\frac{\partial L}{\partial q}\right)-\frac{dp}{dt}\right]\delta q
-\int_{t_1}^{t_2} dt \frac{d}{dt}\left[\delta f -p\delta q \right]
\label{elf}
\end{equation} 
 We  will now invert the relation $C=C(q,\dot q)$ to determine $\dot q=\dot q(q,C)$ and  express $p(q,\dot q)$
in terms of $(q,C)$ obtaining the function $p=p(q,C)$.  In the boundary term in \eq{elf} we treat $f$ as a function of $q$ and $C$, so that the variation of the action can be expressed as:
\begin{eqnarray}
\delta A_C&=&\int_{t_1}^{t_2} dt\left[\left(\frac{\partial L}{\partial q}\right)-\frac{dp}{dt}\right]\delta q+
\left[p(q,C)-\left(\frac{\partial f}{\partial q}\right)_C\right]\delta q\Big|_{t_1}^{t_2}  -\left(\frac{\partial f}{\partial C}\right)_q\delta C\Big|_{t_1}^{t_2} \nonumber\\
&=&\int_{t_1}^{t_2} dt\left[\left(\frac{\partial L}{\partial q}\right)-\frac{dp}{dt}\right]\delta q+
\left[p(q,C)-\left(\frac{\partial f}{\partial q}\right)_C\right]\delta q\Big|_{t_1}^{t_2} 
\end{eqnarray}
since $\delta C=0$ at the end points by assumption. To obtain the same Euler-Lagrange equations,
the second term should vanish for any $\delta q$. This fixes the form of $f$ to be:
\begin{equation}
f(q,C)=\int p(q,C)dq +F(C)
\label{f}
\end{equation}
where the integration is with constant $C$ and $F$ is an arbitrary function. 

Thus, given a lagrangian $L_q(q,\dot q)$ which leads to certain equations of motion when $q$ is held fixed, one can construct a \emph{family} of lagrangians $L_C(q,\dot q,\ddot q)$ which will lead to the {\it same} equations of motion when an arbitrary function $C(q,\dot q)$ is held fixed at the end points. \textit{This family is remarkable in the sense that $L_C$ will be a function of not only
$q,\dot q$, but will also involve $\ddot q$.} In spite of the existence of the $\ddot q$ in the lagrangian, the equations of motion are still of second order in $q$ because of the special structure of the lagrangian. (The results obtained above have an interpretation in terms of canonical transformations etc. which we purposely avoid since we want to stay within the lagrangian framework). So, even though a \textit{general} lagrangian which depends on $\ddot q$ will lead to equations of higher order, there is a host of lagrangians with a \textit{special} structure which will not.

The analysis extends directly to a multicomponent field $q_A(x^a)$ in a spacetime with coordinates $x^a$ where $A$ collectively denotes the tensor indices. Suppose
a lagrangian $ L_q(q_A,\partial q_A)$ gives the field equations when the action is 
 varied  keeping $q_A$ fixed at the boundary $\partial \mathcal{V}$ of a 
spacetime region $\mathcal{V}$. We now want to add to the action a four divergence $\partial_a V^a$
such that the same equations are obtained when the action is varied keeping 
some given functions $U_A^a(q_A,\partial q_A)n_a$  fixed at the boundary  where $n_a$ is the normal to $\partial\mathcal{V}$. As before, we will invert the relation $U_A^a=U_A^a(\partial_a q_A,q_A)$ to determine $\partial_a q_A=\partial_a q_A(q_A,U_A^i)$ and using this will express $\pi^i_A=[\partial L/\partial( \partial_i q^A)]=\pi^i_A
(q_A,\partial_a q_A)$
in terms of $q_A,U_A^i$ getting $\pi^i_A=\pi^i_A(q_A,U_A^j)$. [We are assuming that there are no constraints in the theory and such inversions are possible, for the purpose of illustration.]
Then the lagrangian we are looking for is
\begin{equation}
L_U(\partial^2 q_A,\partial q_A,q_A)=L_q(q_A,\partial q_A)-\partial_a V^a(q_A,\partial q_A)
\end{equation}
with
\begin{equation}
V^j(q_A,U_A^b)=\int \pi^j_A(q_A,U_A^b)dq^A +F^j(U_A^b)
\label{vfix}
\end{equation}
So the same classical field theory can be obtained from a family of lagrangians {\it involving second derivatives } of dynamical variables, provided some arbitrary function of the dynamical variables and their normal derivatives are held fixed at the end points.

When one considers action merely as a tool  to obtain the field equations, the above procedure is acceptable with any $C$ or $U^a_A$. But once
the dynamical variables in the theory have been identified, there are two natural boundary conditions which one may impose on the system. The first one holds $q$ fixed at the end points and the second one keeps the canonical momenta $p$ fixed at the end points. In the second case, $C=p$ and Eq.(\ref{f}) gives $f=qp=q(\partial L/\partial \dot q)$ (ignoring the integration constant);  the corresponding lagrangian is:
\begin{equation}
L_p = L_q - \frac{d}{ dt} \left( q\frac{\partial L_q }{ \partial\dot q} \right) ~.
\label{lbtp}
\end{equation}
The $L_p$ will lead to the same equations of motion when $p(q,\dot q)=(\partial L/\partial \dot q)$ is held
fixed at the end points as can be directly  verified by explicit variation. 
In the hamiltonian language this is summarised by
\begin{equation}
dA=L_qdt-d(qp)=-Hdt +pdq -d(qp)=-Hdt-qdp
\end{equation} 
Since $H$ treats $q,p$ symmetrically, this is just a transformation from $q$ to $p$. 
This result  also has a more natural interpretation in quantum theory: It is easy to show that a path integral defined with $L_p$
will lead to the transition amplitude in momentum space $G(p_2, t_2;p_1, t_1)$, just as a path integral with $L_q$ leads to the transition amplitude in coordinate space$K \left( q_2,t_2;q_1,t_1 \right)$. But our interest is in the existence of higher derivatives of dynamical variables in the lagrangian which is not transparent in the  hamiltonian language and we will continue to use the lagrangian picture.

In the case of field theory, $U_A^a=\pi^a_A$ is independent of $q_A$ and the integral in \eq{vfix} gives $V^j=\pi^j_Aq^A$. So the new lagrangian  is:
\begin{equation}
L_p(\partial^2 q_A,\partial q_A,q_A)=L_q(q_A,\partial q_A)-\partial_i \left[q_A\left(\frac{\partial L_q}{\partial (\partial_i q^A)}\right)\right]
\equiv L_{\rm bulk} + L_{\rm sur}
\end{equation}
It is obvious from this structure (which we shall call, for brevity, the ``$d(qp)$" structure) of the surface term that the surface and bulk terms in the above action are closely related and the knowledge of the surface term will put strong constraints on  $L_{\rm bulk} =L_q$. 
The Einstein-Hilbert action has precisely this form (except for a dimension dependent proportionality constant which becomes unity in D=4 ! See 
\eq{holorel} below).
This is the key to the holography in the action  functionals which we will explore later on.

\section{Actions with Holography}
\label{sec:actholo}

We will now describe a    class of action functionals which allow a decomposition in terms of surface and bulk terms and exhibit a holographic relationship between the two. We will begin by rapidly summarising some of the features of Einstein-Hilbert action in Sec.~\ref{sec:ehaction}  and then generalise them for a wider class in Sec.~\ref{wideaction}.

\subsection{Some features of Einstein-Hilbert action}\label{sec:ehaction}

In this subsection  we will gather together and summarise several results related Einstein-Hilbert action, which we will need later.  
We will not bother to give detailed proofs of these results since we will be providing such proofs --- in a more general context --- in the coming sections. Somewhat longer proofs are presented in Appendix \ref{sec:mainapp} so as not to distract the main discussion.

We begin with the form of the Einstein-Hilbert action for gravity in D-dimensions, given by
\begin{equation}
A_{EH}= \int_\mathcal{V} d^Dx\sqrt{-g}\, L_{EH} =\int_\mathcal{V} d^Dx\sqrt{-g}\, R
\end{equation} 
Using the standard text book expressions for the scalar curvature, one can write the lagrangian in several equivalent forms, all which will be of interest to us later. The simplest one is:
\begin{equation}
L_{EH}\equiv Q_a^{\phantom{a}bcd}R^a_{\phantom{a}bcd}; \qquad Q_a^{\phantom{a}bcd}=\frac{1}{2}(\delta^c_ag^{bd}-\delta^d_ag^{bc})
\label{lisrq}
\end{equation}
The tensor  $Q_a^{\phantom{a}bcd}$ is the only fourth rank tensor that can be constructed from the metric (alone) that has all the symmetries of the curvature tensor. In addition it has zero divergence on all indices, $\nabla_a Q^{abcd}=0$ etc. Since the curvature tensor $R^a_{\phantom{a}bcd}$ can be expressed entirely in terms of $\Gamma^i_{kl}$ and $\partial_j\Gamma^i_{kl}$ without requiring  $g^{ab}$,  there is a nice separation between dependence on the metric (through 
$Q_a^{\phantom{a}bcd}$ alone) and dependence on connection and its derivative through $R^a_{\phantom{a}bcd}$. This separation is useful in certain variational calculations when we treat them as independent. With $g^{ab},\Gamma^i_{kl},R^a_{\phantom{a}bcd}$ (instead of $g_{ab}$ and its first and second derivatives) treated as independent variables, the vacuum Einstein's equations take a very simple form:
\begin{equation}
\left(\frac{\partial \sqrt{-g}L_{EH}}{\partial g^{ab}}\right)=
R^a_{\phantom{a}bcd}\left(\frac{\partial \sqrt{-g}Q_a^{\phantom{a}bcd}}{\partial g^{ab}}\right)=0
\label{ordder}
\end{equation} 
That is, Einstein's equations arise through ordinary partial differentiation of the Lagrangian density with respect to $g^{ab}$, keeping $\Gamma^i_{kl}$ and $\partial_j\Gamma^i_{kl}$ as constant.

If we raise one index of the curvature tensor, the Einstein-Hilbert lagrangian can be written in another interesting form as
\begin{equation}
L_{EH}\equiv \delta_{ab}^{cd}R^{ab}_{cd}; \qquad \delta_{ab}^{cd}=\frac{1}{2}(\delta^c_a\delta_b^{d}-\delta^d_a\delta_b^{c})
\end{equation}
where $\delta_{ab}^{cd}$ is the alternating (`determinant') tensor. The importance of this form lies in the fact that it allows the  generalisation \cite{love} to a  lagrangian containing a product of, say, $m$ curvature tensors, which --- as we shall see ---  will share many properties of Einstein-Hilbert action. 

We will now turn to more nontrivial  aspects of the Einstein-Hilbert action which  provides the key motivation to this work. Since $L_{EH}$  is linear in second derivatives of the metric, it is clear that $\sqrt{-g}L_{EH}$ can be written as a sum $L_{bulk}+L_{sur}$ where $L_{bulk}$ is quadratic in the first derivatives of the metric and
$ L_{sur}$ is a total derivative which leads to a surface term in the action. From \eq{lisrq}
it is easy to obtain this separation as
 \begin{equation}
\sqrt{-g}L_{\rm EH}=2\partial_c\left[\sqrt{-g}Q_a^{\phantom{a}bcd}\Gamma^a_{bd}\right]
+2\sqrt{-g}Q_a^{\phantom{a}bcd}\Gamma^a_{dk}\Gamma^k_{bc}\equiv L_{\rm sur} + L_{\rm bulk} 
\label{gensq}
 \end{equation} 
with
\begin{equation}
L_{\rm bulk}=2\sqrt{-g}Q_a^{\phantom{a}bcd}\Gamma^a_{dk}\Gamma^k_{bc};\qquad
L_{\rm sur}=2\partial_c\left[\sqrt{-g}Q_a^{\phantom{a}bcd}\Gamma^a_{bd}\right]
\equiv\partial_c\left[\sqrt{-g}V^c\right]
\label{sep}
\end{equation}  
where the last equality defines the D-component object $V^c$, which --- of course --- is not a vector. (The proof is given in Appendix \ref{sec:appone}.) Even in this form, the metric dependence is confined to  $Q_a^{\phantom{a}bcd}$.
As is well known, one can obtain Einstein's equations varying only $L_{bulk}$  keeping $g_{ab}$ fixed at the boundary.

The first nontrivial result regarding Einstein-Hilbert action is a simple relation \cite{tpreview} between $L_{\rm bulk}$
and $L_{\rm sur}$ allowing $L_{\rm sur}$ to be determined completely by $L_{\rm bulk}$. 
(As discussed in Sec. \ref{sec:intro}, we call such a relation  holographic).  Using $g_{ab}$ and $\partial_cg_{ab}$ as the independent variables in $L_{\rm bulk}$ one can prove that:
 \begin{equation}
L_{\rm sur}=-\frac{1}{[(D/2)-1]}\partial_i\left(g_{ab}\frac{\partial L_{\rm bulk}}{\partial (\partial_ig_{ab})}\right)
\label{holorel}
\end{equation}
The ``$d(qp)$'' structure of $L_{\rm sur}$  suggests that $L_{\rm EH}$ is obtained
from $L_{\rm bulk}$ by a transformation from coordinate space to momentum space, as has been noticed before in literature \cite{tpreview}. One of our aims will be to obtain a suitable generalisation of this result to a wider class of lagrangians. We will prove a more general result, viz. \eq{result1} below, of which \eq{holorel} is a special case.

We note, in passing, that there are other ways of stating the holographic relation. For example, we can write a relation of the form:
\begin{equation}
L_{\rm sur} = -\partial_p \left(\delta^q_r \frac{\partial L_{\rm bulk}}{\partial \Gamma^q_{pr}}
\right)
\label{holo}
\end{equation} 
The proof is straight forward  but there is one caveat.
In evaluating the partial derivative on the right hand side we hold $Q_a^{\phantom{a}bcd}$
fixed and treat all components of $\Gamma^k_{bc}$ as independent variables with no symmetry requirements; that is, we take
$(\partial \Gamma^a_{bc}/\partial \Gamma^i_{jk})=\delta^a_i\delta^j_b\delta^k_c$ to obtain:
\begin{equation}
\delta^q_r\frac{\partial L_{bulk}}{\partial \Gamma^q_{pr}}=\delta^q_r\left[2\sqrt{-g}
\left(Q_a^{\phantom{a}prd}\Gamma^a_{dq}+Q_q^{\phantom{q}uvp}\Gamma^r_{uv}\right)\right]
=2\sqrt{-g}
\left(Q_a^{\phantom{a}prd}\Gamma^a_{dr}+Q_r^{\phantom{q}uvp}\Gamma^r_{uv}\right)
\label{dldG}
\end{equation}
Obviously, the order of lower indices in $\Gamma^a_{bc}$ in \eq{sep} is important in arriving at this result.   
After the derivative is computed we will impose the condition that the $\Gamma^k_{bc}$ are related to the metric by the standard relation. Then, the symmetry of $\Gamma^a_{dr}$ in $d,r$ makes the first term vanish (since $Q_a^{\phantom{a}prd}= - Q_a^{\phantom{a}pdr}$) and the result becomes
\begin{equation}
\delta^q_r\frac{\partial L_{bulk}}{\partial \Gamma^q_{pr}}=2\sqrt{-g}
\left(Q_r^{\phantom{q}uvp}\Gamma^r_{uv}\right)=-2\sqrt{-g}
\left(Q_r^{\phantom{q}upv}\Gamma^r_{uv}\right)
\end{equation} 
A comparison with the definition of $L_{\rm sur}$ in \eq{sep} leads to \eq{holo}. 

In the same manner we can also prove the following results \cite{tpconf} which determine the bulk and total lagrangians in terms of the surface term (which is probably truer to the spirit of the term holography):
\begin{equation}
L=\frac{1}{2}R^a_{\phantom{a}bcd}\left(\frac{\partial V^c}{\partial \Gamma^a_{bd}}\right);
\quad
L_{bulk}=\sqrt{-g}\left(\frac{\partial V^c}{\partial \Gamma^a_{bd}}\right)
\Gamma^a_{dk}\Gamma^k_{bc}
\label{realholo}
\end{equation}  
Thus the knowledge of the functional form of $L_{sur}$ or --- equivalently --- that of $V^c$
allows us to determine $L_{bulk}$ and even $L$.
The first relation also shows that  $(\partial V^c/\partial \Gamma^a_{bd})$ is generally covariant in spite of the appearance.

The fact that one needs to first treat $\Gamma^a_{bc}$ as independent and then impose the metric compatibility makes the above results less attractive than the one stated in \eq{holorel}. \textit{On the other hand, \eq{holo} and \eq{realholo} do not use
 the explicit form of  $Q_a^{\phantom{a}bcd}$.} In the case Einstein-Hilbert action,  $Q_a^{\phantom{a}bcd}$ is independent of curvature and depends only on metric but in the next section we will consider $Q_a^{\phantom{a}bcd}$ that is made from metric, curvature tensor and possibly covariant derivatives of the curvature tensor --- all of which can be held fixed while varying $\Gamma^a_{bc}$, if we choose the metric, the curvature tensor,its covariant derivatives and also $\Gamma^a_{bc}$ as independent variables.  All such lagrangians of the form in \eq{gensq} will satisfy \eq{holo} and \eq{realholo}. In contrast, \eq{holorel} uses the specific form of $Q_a^{\phantom{a}bcd}$
 given in \eq{lisrq}
(A straight forward proof of \eq{holorel}, starting from \eq{holo} and changing variables is given in the Appendix \ref{sec:apptwo}, thereby connecting up the two and demonstrating where \eq{lisrq} is used.)

Before we conclude this section, we would like to comment on a few other issues related to the surface term. The separation of Einstein-Hilbert action into surface and bulk terms in \eq{gensq}
is a standard text book result. While neither term is generally covariant, the \textit{variation} of either term  with respect to metric is generally covariant (when the metric is held fixed at the boundary) so that, for example, $L_{bulk}$ can lead to the standard field equations when the metric is held fixed at the boundary. It is, of course, possible to add other surface terms to Einstein-Hilbert action so that the same field equations are obtained under variation of the metric, when the metric is held fixed at the boundary. The most popular one is the Gibbons-Hawking term $A_{GH}$ which is the integral over the trace of the extrinsic curvature of the boundary \cite{gh}. The $A_{sur}$ in \eq{gensq} is not equal to $A_{GH}$ in general but matches under a particular coordinate choice (See the Appendix of \cite{tpreview} for a discussion; of course, the variations of the two terms always match). In the interpretation of \eq{gensq} as having the ``$d(qp)$ structure", we have treated all components of $g_{ab}$ at the same footing in the spirit of lagrangian formulation. It is well known from the Hamiltonian structure of the theory that $g_{00}$ and $g_{0\alpha}$ 
are constraint variables in the sense that their time derivatives do not occur in the lagrangian. If we integrate the $L_{sur}$ over a volume bounded by two spacelike surfaces $\Sigma_{1,2}$, we will pick up $[g_{\mu\nu}\partial L_{bulk}/\partial(\partial_0 g_{\mu\nu})\equiv g_{\mu\nu}\pi^{\mu\nu}]$ involving only the correct dynamical variables $g_{\mu\nu}$ and their canonical momenta $\pi^{\mu\nu}=K^{\mu\nu}-g^{\mu\nu}K$ on these surfaces. If we further choose the gauge $g_{0\mu}=0$, then we obtain the integral over
$g_{\mu\nu}\pi^{\mu\nu}=-2K$, in the standard $D=4$ case, which is the same as $A_{GH}$. So the claim
that \eq{gensq} has the ``$d(qp)$ structure" is quite appropriate even from this perspective.

The $A_{GH}$ has a formal general covariance (which our term lacks) but this is obtained at the cost of foliation dependence. The relation between foliation dependence and general covariance is worth emphasizing: One would have considered a component
of a tensor, say, $T_{00}$ as not generally covariant. But a quantity $\rho=T_{ab}u^au^b$ is a generally covariant scalar which will reduce to $T_{00}$ in a local frame in which $u^a=(1,0,0,0)$.
It is appropriate to say that $\rho$ is generally covariant but foliation dependent. In fact, any term which is not generally covariant can be recast in  a generally covariant form by introducing a foliation dependence. The $A_{GH}$ uses the normal vector $n^i$ of the boundary in a similar manner. But since our boundary term will reduce to $A_{GH}$ under a particular coordinate choice, all the results which we quote will similarly applicable, under this coordinate choice, to
$A_{GH}$ as well. The situation becomes more complicated in the case of  general lagrangians and we will comment on this later.

\subsection{Actions with Holography:  Generalisation}\label{wideaction}

The Einstein-Hilbert action is usually introduced by using the fact that it is the only generally covariant scalar that can be built from metric and its derivatives and is linear in the second derivatives. This guarantees that the variational principle could be made to work, albeit with some  unusual boundary conditions. It is, therefore, quite surprising that the action possesses several \textit{other} peculiar properties, in particular the holographic relations between the surface and bulk terms.

On the other hand, it is quite possible that Einstein-Hilbert action describes the  low energy limit of an effective theory and $L_{\rm EH}$ is just a first term in a series of terms which will involve other scalars (like $R^2,R_{ab}R^{ab}$) that can be constructed from the metric and curvature. Several possible choices exist for such low energy effective action all of which are consistent with the diffeomorphism invariance of the low energy theory.
Any extra symmetry, like the holographic relation, could then serve as a powerful guiding principle in constraining or determining the higher order corrections to the action principle. This leads us to ask:  What is the most general action for gravity which satisfies holographic conditions~?  We will now address this question.

Since the relations in \eq{holo} and \eq{realholo} are linear in the lagrangian, it follows that if two lagrangians individually satisfy these relations, so will their sum with arbitrary constant coefficients.
This allows us to investigate the individual terms in a sum of terms separately and also allows us ignore relative coupling constants between them.
Further, since our derivation of \eq{holo} (or \eq{realholo}) did not use the explicit form of $Q_a^{\phantom{a}bcd}$ we already know that any lagrangian of the form 
\begin{equation}
\sqrt{-g}L=2\partial_c\left[\sqrt{-g}Q_a^{\phantom{a}bcd}\Gamma^a_{bd}\right]
+2\sqrt{-g}Q_a^{\phantom{a}bcd}\Gamma^a_{dk}\Gamma^k_{bc}\equiv L_{\rm sur} + L_{\rm bulk} 
\label{gensq1}
 \end{equation}
will satisfy the  relations in \eq{holo} and \eq{realholo}
provided: (i) $Q^{abcd}$ has all the symmetries of curvature tensor and (ii) one can keep $Q^{abcd}$ constant while differentiating with respect to $\Gamma^a_{bc}$ treating all the components as independent. 

Let us now consider a general lagrangian of the form in \eq{gensq1} with
$Q^{abcd}=Q^{abcd}(g^{ab},R^a_{\phantom{a}bcd},\nabla_jR^a_{\phantom{a}bcd}....)$ depending on the metric, curvature tensor and its covariant derivatives. We will follow the standard principle that, when varying an action, the dynamical variable $q_A$ and each of higher derivatives $\partial q_A, \partial^2 q_A .....$etc are to be treated as independent. If a lagrangian $L$ depends on the  metric $g^{ab}$, curvature $R^a_{\phantom{a}bcd}$ and its derivatives, the dynamical variable and its derivatives are the set $[g^{ab},\partial_cg^{ab},
\partial_d\partial_cg^{ab},....$] and we treat them as independent. 
Instead of treating $[g^{ab},\partial_cg^{ab},
\partial_d\partial_cg^{ab},....$] as the independent variables,  it is convenient to use $[g^{ab},
\Gamma^i_{kl},R^a_{\phantom{a}bcd},...]$ as the independent variables
and we trade off the
second (and higher) derivatives of the metric $[\partial_d\partial_cg^{ab},....]$, in favour of 
curvature tensor and its derivatives \cite{ps2}.
The curvature tensor $R^a_{\phantom{a}bcd}$ can be expressed entirely in terms of $\Gamma^i_{kl}$ and $\partial_j\Gamma^i_{kl}$ and is \textit{independent} of $g^{ab}$.
 Then, we can indeed keep
$R^a_{\phantom{a}bcd}$ and its derivatives (as well as the metric itself) constant while differentiating with respect $\partial_ig_{kl}$. 
Therefore, the lagrangian in \eq{gensq1} with $Q^{abcd}$ being a tensor with the symmetries of curvature tensor, constructed from
metric, curvature and covariant derivatives of the curvature will satisfy \eq{holo} and \eq{realholo}. 

But, in general, the expression in \eq{gensq1} will \textit{not} be a generally covariant scalar
since it is expressed in terms of $\Gamma^a_{bc}$ etc. We need to ascertain the condition on $Q^{abcd}$ such that the lagrangian is generally covariant. This turns out to be surprisingly easy and insightful. By straight forward algebra, one can prove (see Appendix \ref{sec:appone}) the following identity:
\begin{equation}
\sqrt{-g}L=2\partial_c\left[\sqrt{-g}Q_a^{\phantom{a}bcd}\Gamma^a_{bd}\right]
+2\sqrt{-g}Q_a^{\phantom{a}bcd}\Gamma^a_{dj}\Gamma^j_{bc}
 = \sqrt{-g}Q_a^{\phantom{a}bcd}R^a_{\phantom{a}bcd}+2\sqrt{-g}\Gamma^a_{bd}\nabla_cQ_a^{\phantom{a}bcd}
 \label{simple}
\end{equation} 
Obviously, general covariance only requires the condition $\nabla_cQ_a^{\phantom{a}bcd}=0$. Because of the symmetries of the $Q_a^{\phantom{a}bcd}$ its divergence on any of the indices vanishes. Thus, we shall hereafter consider lagrangians of the form:
\begin{equation}
\sqrt{-g}L=\sqrt{-g}Q_a^{\phantom{a}bcd}R^a_{\phantom{a}bcd}; \qquad \nabla_cQ_a^{\phantom{a}bcd}=0
\label{basicl}
\end{equation}  
We have already proved that all such generally covariant lagrangians
are holographic; i.e., they allow a separation into bulk and surface terms which are related by \eq{holo} and \eq{realholo}. 

The simplicity of this result suggests that there could be a more geometric way of interpreting it. This is indeed true \cite{tpconf}.
We know that the one can write the curvature tensor in terms of the two form by
 $\mathcal{R}^a_{\phantom{a}b} = (1/2!) R^a_{\phantom{a}bcd} \, w^c \wedge \, w^d$ where $w^a$ are the  basis one forms. Similarly one can
 introduce a  two form for $Q_{abcd}$ with $\mathcal{Q}^a_{\phantom{a}b} =(1/2!) Q^a_{\phantom{a}bcd}\, w^c \wedge w^d$. Further,
 using $\mathcal{R}^a_{\phantom{a}b} =  d\,   \Gamma^a_{\phantom{a}b} + \Gamma^a_{\phantom{a}c}\, \wedge \, \Gamma^c_{\phantom{c}b}$ where $\Gamma^a_{\phantom{a}b}$ are the curvature forms,
 we can write the 
 Lagrangian in \eq{basicl} as
 \begin{equation}
L = \ast \mathcal{Q}^a_{\phantom{a}b}\wedge \mathcal{R}^b_{\phantom{b}a} 
= \ast \mathcal{Q}^a_{\phantom{a}b}\wedge \left( d \Gamma^b_{\phantom{b}a} +  \Gamma^b_{\phantom{b}c}\wedge
\Gamma^c_{\phantom{c}a}\right) 
= d\left( \ast \mathcal{Q}^a_{\phantom{a}b}\wedge \Gamma^b_{\phantom{b}a}
\right) + \ast\mathcal{Q}^a_{\phantom{a}b}\wedge \Gamma^b_{\phantom{b}c}\wedge \Gamma^c_{\phantom{c}a}
\end{equation}
provided the $\mathcal{Q}^a_{\phantom{a}b}$ satisfies the condition:
$d\left( \ast \mathcal{Q}^a_{\phantom{a}b}\right) =0$ corresponding to $\nabla_c Q_a^{\phantom{a}bcd} =0$. The separation between bulk and surface terms, just as in the case of \eq{gensq1}, is obvious.

While discussing the corresponding situation in the case of Einstein-Hilbert action we commented on the, alternative, Gibbons-Hawking surface term $A_{GH}$.
The situation becomes more complicated when we move to more general lagrangians. The analogue of 
$A_{GH}$ for more general lagrangians is difficult to come by and --- as far as the authors know --- there is no algorithmic procedure for finding them. The expressions given in literature for even the Gauss-Bonnet case \cite{gbboundary} are fairly complicated and their physical meanings are unclear. But our $L_{sur}$ is well defined for a wide class of lagrangians
and possesses some of the attractive properties, which is encouraging. The lack of manifest general covariance is not of much concern since this issue exists even for Einstein-Hilbert action. (In specific cases, like in asymptotically flat spacetimes possessing a horizon, the surface term actually turns out to be generally covariant and gives the horizon entropy; see Sec. \ref{sec:entro}). 

The structure of the theory is thus specified by a single divergence-free fourth rank tensor $Q_a^{\phantom{a}bcd}$ having the symmetries of the curvature tensor. If we think of gravity as low energy effective theory, the semi classical,
 action for gravity can now be determined from the derivative expansion
of $Q_a^{\phantom{a}bcd}$ in powers of number of derivatives: 
\begin{equation}
Q_a^{\phantom{a}bcd} (g,R) = \overset{(0)}{Q}_a{}^{bcd} (g) + \alpha\, \overset{(1)}{Q}_a{}^{bcd} (g,R) + \beta\, \overset{(2)}{Q}_a{}^{bcd} (g,R^2,\nabla R) + \cdots
\label{derexp}
\end{equation} 
where $\alpha, \beta, \cdots$ are coupling constants. At the lowest order, $Q_a^{\phantom{a}bcd}$
has to be built from just the metric and  next order correction will have $Q_a^{\phantom{a}bcd}$ depending on $R^a_{\phantom{a}bcd}$ linearly as well as on the metric etc.

Interestingly enough, the condition  $\nabla_cQ_a^{\phantom{a}bcd}=0$ encompasses \textit{all} the gravitational theories (in D dimensions) in which
the field equations are no higher than second degree, though we did \textit{not }demand that explicitly \cite{tpconf}. To see this, let us consider the possible fourth rank tensors $Q^{abcd}$ which (i) have the symmetries of curvature tensor; (b) are divergence-free; (iii) made from $g^{ab}$ and  $R^a_{\phantom{a}bcd}$. If we do not use the curvature tensor, then we have just one choice
 made from metric given in \eq{lisrq} and will lead
 to the Einstein-Hilbert action.
Next, if we allow for $Q_a^{\phantom{a}bcd}$ to depend linearly on curvature, then we have the following 
additional choice of  tensor with required symmetries:
\begin{equation}
Q^{abcd}=R^{abcd} -  G^{ac}g^{bd}+ G^{bc}g^{ad} + R^{ad}g^{bc} - R^{bd}g^{ac} 
\label{ping}
\end{equation} 
In four dimensions, this tensor is essentially the double-dual of
$R_{abcd}$ and in any dimension can be obtained from $R_{abcd}$ using the alternating tensor \cite{gliner} we get
\begin{equation}
L=\frac{1}{2}\left(g_{ia}g^{bj}g^{ck}g^{dl}-4g_{ia}g^{bd}g^{ck}g^{jl}
+\delta^c_a\delta^k_ig^{bd}g^{jl}\right)R^i_{\phantom{i}jkl}R^a_{\phantom{a}bcd}
=\frac{1}{2}\left[R^{abcd}R_{abcd}-4R^{ab}R_{ab}+R^2\right]
\end{equation} 
This just the Gauss-Bonnet action which is a pure divergence in four dimensions but not in higher dimensions.
The unified procedure for deriving Einstein-Hilbert action and Gauss-Bonnet  action [essentially from the holographic condition  and $\nabla_cQ_a^{\phantom{a}bcd}=0$] shows that they are more closely related  to each other than previously suspected. The fact that \textit{several string theoretical models get Gauss-Bonnet  type terms as corrections}, after appropriate field redefinitions, \cite{zw} is noteworthy in this regard. 

Further, both  Einstein-Hilbert lagrangian and  Gauss-Bonnet  lagrangian can be written in a condensed notation using alternating tensors as: 
\begin{equation}
L_{EH}=\delta^{13}_{24}R^{24}_{13};\qquad 
L_{GB}=\delta^{1357}_{2468}R^{24}_{13}R^{68}_{57}
\end{equation} 
where the numeral $n$ actually stands for an index $a_n$ etc. The obvious generalisation leads to the Lanczos-Lovelock lagrangian \cite{love}:
\begin{equation}
L_m=\delta^{1357...2k-1}_{2468...2k}R^{24}_{13}R^{68}_{57}
....R^{2k-2\,2k}_{2k-3\,2k-1}; \qquad k=2m
\label{lll}
\end{equation} 
where $k=2m$ is an even number. The $L_m$ is clearly a homogeneous function of degree $m$ in
the curvature tensor $R^{ab}_{cd}$ so that it  can also  be expressed in the form:
\begin{equation}
L=\frac{1}{m}\left(\frac{\partial L}{\partial R^a_{\phantom{a}bcd} }\right)R^a_{\phantom{a}bcd}\equiv \frac{1}{m}P_a^{\phantom{a}bcd}R^a_{\phantom{a}bcd}.
\end{equation} 
where we have defined $P_a^{\phantom{a}bcd}\equiv (\partial L/\partial R^a_{\phantom{a}bcd} )$
so that $P^{abcd}=mQ^{abcd}$. It can be directly verified that for these lagrangians (see Appendix \ref{sec:appadded}):
\begin{equation}
\nabla_cP^{ijcd}=0
\label{condition}
\end{equation} 
Because of the symmetries,  $P^{abcd}$ is divergence-free in \textit{all} indices.
These lagrangians, therefore, belong to the class described by \eq{basicl} and --- more importantly for our purpose --- they allow a separation in to bulk and surface terms as given by \eq{gensq1} with the two parts satisfying \eq{holo} and \eq{realholo}.
It is also possible to prove the following result for these lagrangians:  If we treat $\Gamma^a_{bc}$ as independent of $g^{ab}$ and vary it, keeping 
$g^{ab}$ fixed, then it is easy to show, [using the results of Appendix \ref{sec:appfour}] that
$\delta L/\delta \Gamma^a_{bc}=0$ if $\nabla_aP^{abcd}=0$. So this condition allows one to vary
$\Gamma^a_{bc}$s  independently of $g^{ab}$ as in Palatini formulation of general relativity, though we will not use this result.
The $m=1$ and $m=2$ give Einstein-Hilbert and Gauss-Bonnet lagrangians.
We shall now prove a host of relations for this class of lagrangians.

The first result  is that, the equations of motion for these lagrangians take a particularly simple form. To see this, let us consider a general action of the form
\begin{equation}
A= \int_\mathcal{V} d^Dx\sqrt{-g}\, L[g^{ab},R^a_{\phantom{a}bcd}] 
\end{equation} 
in which we have ignored higher derivatives of $R^a_{\phantom{a}bcd}$ for simplicity. The variation of the action
can be easily computed to give the result (see Appendix \ref{sec:appthree} for details)
\begin{equation}
\delta A = \delta \int_\mathcal{V} d^Dx\sqrt{-g}\, L = \int_\mathcal{V} d^Dx \, \sqrt{-g} \, E_{ab} \delta g^{ab} + 
\int_\mathcal{V} d^Dx \, \sqrt{-g} \, \nabla_j \delta v^j
\label{deltaa}
\end{equation}
where
\begin{equation}
\sqrt{-g}E_{ab}\equiv\left( \frac{\partial \sqrt{-g}L}{\partial g^{ab}} 
-2\sqrt{-g}\nabla^m\nabla^n P_{amnb} \right); \qquad P_a^{\phantom{a}bcd}\equiv (\partial L/\partial R^a_{\phantom{a}bcd} )
\label{eab}
\end{equation} 
and
\begin{equation}
\delta v^j \equiv [2P^{ibjd}(\nabla_b\delta g_{di})-2\delta g_{di}(\nabla_cP^{ijcd})]
\label{genvc}
\end{equation} 
This result is completely general. We now see that the equations of motion simplify significantly
for a subclass of lagrangians which satisfy \eq{condition} and is given by
\begin{equation}
\frac{\partial \sqrt{-g}L}{\partial g^{ab}}=0
\label{ordder1}
\end{equation} 
That is, just setting the \textit{ordinary derivative} of lagrangian density with respect to $g^{ab}$ to zero will give the equations of motion, as in the case Einstein-Hilbert action.

It also follows that, for the $m$th  Lanczos-Lovelock lagrangian, $L_m$ [given by \eq{lll}], the trace of the equations of motion is proportional to the lagrangian:
\begin{equation}
g^{ab}E_{ab}=g^{ab}\frac{\partial \sqrt{-g}L_m}{\partial g^{ab}}=-[(D/2)-m]\sqrt{-g}L_m;\qquad
g_{ab}E^{ab}=g_{ab}\frac{\partial \sqrt{-g}L_m}{\partial g_{ab}}=[(D/2)-m]\sqrt{-g}L_m
\label{treom}
\end{equation} 
This \textit{off-shell} relation is easy to prove from the fact that we need to introduce $m$ factors of $g^{ab}$ in  \eq{lll} to proceed from $R^a_{\phantom{a}bcd}$ to $R^{ab}_{cd}$ and that $\sqrt{-g}$ is homogeneous function of $g^{ab}$ of degree $-D/2$.
Further, we can prove that (see Appendix \ref{sec:appfour} for the proof) for \textit{any} lagrangian:
\begin{equation}
g_{ab} \frac{\delta L}{\delta (\partial_i g_{ab})}=-2\sqrt{-g}\left(\frac{\partial L}{\partial R_{nbid} }\right)\Gamma_{nbd}=-2\sqrt{-g}P^{nbid} \Gamma_{nbd}
\label{eulerder}
\end{equation} 
where the \textit{Euler derivative} is defined as 
\begin{equation}
\frac{\delta K[\phi,\partial_i\phi,...]}{\delta\phi}=
\frac{\partial K[\phi,\partial_i\phi,...]}{\partial\phi}-
\partial_a \left[\frac{\partial K[\phi,\partial_i\phi,...]}{\partial(\partial_a\phi)}\right]+\cdots
\end{equation}
In the case of  Lanczos-Lovelock lagrangians, $P^{nbid}=mQ^{nbid}$ so that we get the relation:
\begin{equation}
\partial_i\left[g_{ab} \frac{\delta L}{\delta (\partial_i g_{ab})}\right]
=-mL_{sur}
\label{mlsur}
\end{equation} 
This shows that $m$ times the surface term is indeed of the  $``d(qp)"$ structure provided the momentum is defined using the \textit{total} lagrangian $L$ and Euler derivative. 
   We also know that  all lagrangians of the form in \eq{gensq1} satisfies \eq{holo} and \eq{realholo} as well, with  a specific prescription for evaluation of the derivative.
   Thus we have established three different holographic relations for  Lanczos-Lovelock lagrangians.

  Since the Einstein-Hilbert lagrangian corresponds to  Lanczos-Lovelock lagrangian with $m=1$,   \eq{mlsur} is valid for $L_{\rm EH}$ as well. But the relation in \eq{mlsur} should be distinguished from
 \eq{holorel} which shows that a similar relation also holds with \textit{bulk} lagrangian $L_{\rm bulk}$ rather than with \textit{total} lagrangian $L_{\rm EH}$.
We shall now take up the generalisation of the relation \eq{holorel} (between $L_{bulk}$ and $L_{sur}$) for the  Lanczos-Lovelock case  when $Q_a^{\phantom{a}bcd}$
depends on the metric as well as the curvature. (The result has a direct generalisation even for some cases in which $Q_a^{\phantom{a}bcd}$ depends on the derivatives of the curvature tensor as well \cite{fgi}; however, to keep the argument transparent, we will discuss the simpler case, which --- in any case --- is more relevant to us.). We will prove that:
\begin{equation}
[(D/2) - m]L_{sur} =-\partial_i \left[ g_{ab} \frac{\delta L_{bulk}}{\delta (\partial_i g_{ab})}
  +\partial_jg_{ab} \frac{\partial L_{bulk}}{\partial (\partial_i \partial_jg_{ab})}
  \right] 
\label{result1} 
\end{equation} 
 Before we give the proof, we will make couple of comments on the result. First, in the case of Einstein-Hilbert lagrangian, the $L_{bulk}$ does not involve the second derivatives of the metric. Therefore, the second term in the right hand side of \eq{result1} is absent and --- in the first term --- we can replace the Euler derivative by ordinary partial derivative. This leads to \eq{holorel} as it should. Second, the terms in the right hand side, for the general case, can be thought of as one form of  generalisation of  $``d(qp)"$ for theories with higher derivatives. 

The proof of \eq{result1} is based on a simple homology argument and combinatorics, generalising a corresponding proof for \eq{holorel} in Einstein-Hilbert case (given in the Appendix of ref.\cite{tpreview}).
Consider any Lagrangian $L(g_{ab}, \partial_i g_{ab}, \partial_i\partial_j g_{ab})$
and let $E^{ab}[L]$ denote the Euler-Lagrange function resulting from $L$: 
  \begin{equation}
  E^{ab}[L] \equiv \frac{\partial L}{\partial g_{ab}} - \partial_i \left[\frac{\partial L}{\partial(\partial_i g_{ab})} \right]+\partial_i \partial_j\left[\frac{\partial L}{\partial(\partial_i\partial_j g_{ab})} \right]
  \end{equation}
  Forming the contraction $g_{ab} E^{ab}$ and manipulating the terms, we can rewrite this equation as:
  \begin{equation}
  g_{ab} E^{ab}[L] = g_{ab}\frac{\partial L}{\partial g_{ab}} +  
  (\partial_i g_{ab}) \frac{\partial L}{\partial (\partial_i g_{ab})} +
  (\partial_i \partial_j g_{ab}) \frac{\partial L}{\partial (\partial_i \partial_j g_{ab})}
  - \partial_i \left[ g_{ab} \frac{\delta L}{\delta (\partial_i g_{ab})}
  +\partial_jg_{ab} \frac{\partial L}{\partial (\partial_i \partial_jg_{ab})}
  \right] 
  \label{qF}
  \end{equation}
We will now apply this relation to the bulk lagrangian $L_{bulk}^{(m)}= 2\sqrt{-g}Q_a^{\phantom{a}bcd}\Gamma^a_{dj}\Gamma^j_{bc}$ corresponding to the $m$ th order  Lanczos-Lovelock lagrangian. (Hereafter, we will simplify notation by just calling it $L_{bulk}$; it is understood that we are dealing with the $m$th order  Lanczos-Lovelock lagrangian throughout.) Since both $L_m$ and $L_{bulk}$  lead to the same equations of motion,  $E^{ab}[L_m]=E^{ab}[L_{bulk}]$. Hence, using \eq{treom}, we find the left hand side of \eq{qF} to be $[(D/2)-m]\sqrt{-g}L_m$. We will next show that the first three terms in the right hand side add up to give $[(D/2)-m]L_{bulk}$. Bringing this term to the left hand side and using $L_{sur}=\sqrt{-g}L-L_{bulk}$ will then lead to \eq{result1}. 

To prove this, let us write  $L_{bulk}/\sqrt{-g}$ entirely in terms of $g^{ab},\partial_ig_{ab}$ and
$\partial_j\partial_ig_{ab}$ by multiplying out completely. In any given term, let us assume
there are $n_0$ factors of $g^{ab}$, $n_1$ factors of $\partial_ig_{ab}$ and $k$ factors of $\partial_i\partial_jg_{ab}$. Then homogeneity
implies that for this particular term (labelled by $k$, which is the number of
$\partial_i\partial_jg_{ab}$, that occur in it), the first three terms in the right hand side of
\eq{qF} are given by
\begin{equation}
g_{ab}\frac{\partial L_{bulk}^{(k)}}{\partial g_{ab}}=[(D/2)-n_0]L_{bulk}^{(k)};\quad
(\partial_i g_{ab}) \frac{\partial L_{bulk}^{(k)}}{\partial (\partial_i g_{ab})}=n_1L_{bulk}^{(k)};\quad
(\partial_i \partial_j g_{ab}) \frac{\partial L_{bulk}^{(k)} }{\partial (\partial_i \partial_j g_{ab})}
=kL_{bulk}^{(k)}
\label{firststep}
\end{equation}
(In the first relation $D/2$ comes from the $\sqrt{-g}$ factor and the sign flip on $n_0$ is because of switching over from $g^{ab}$ to $g_{ab}$.). 
Since all the indices --- the 2 upper indices from each $g^{ab}$, 3 lower indices from each $\partial_ig_{ab}$, 4 lower indices from each $\partial_j\partial_ig_{ab}$ ---  are to be contracted out, we must have $2n_0=3n_1+4k$ which fixes $n_0$ in terms of $n_1$ and $k$. We next note that, $Q_a^{\phantom{a}bcd}$
is made of $(m-1)$ factors of curvature tensor and each curvature tensor has the structure
$R\simeq [\partial^2g + (\partial g)^2]$. If we multiply out $(m-1)$ curvature tensors, a generic term in the product will have $k$ factors of $\partial^2g$ and $(m-1-k)$ factors of $(\partial g)^2$. In addition,
the two $\Gamma$'s in $L_{bulk}\simeq Q\Gamma\Gamma$ will contribute two more factors of 
$(\partial g)$. So, for this generic term, $n_1=2(m-1-k)+2=2(m-k)$. Using our relation
$2n_0=3n_1+4k$, we find $n_0=3m-k$. Substituting into \eq{firststep} we get
\begin{equation}
g_{ab}\frac{\partial L_{bulk}^{(k)}}{\partial g_{ab}}=[(D/2)-3m+k]L_{bulk}^{(k)};\quad
(\partial_i g_{ab}) \frac{\partial L_{bulk}^{(k)}}{\partial (\partial_i g_{ab})}=2(m-k)L_{bulk}^{(k)};\quad
(\partial_i \partial_j g_{ab}) \frac{\partial L_{bulk}^{(k)} }{\partial (\partial_i \partial_j g_{ab})}
=kL_{bulk}^{(k)}
\end{equation}
 Though each of these terms depends on $k$, the sum of the three terms is independent of $k$  leading to the same contribution from each term. So we get:
\begin{equation}
g_{ab}\frac{\partial L_{bulk}}{\partial g_{ab}}+
(\partial_i g_{ab}) \frac{\partial L_{bulk}}{\partial (\partial_i g_{ab})}+
(\partial_i \partial_j g_{ab}) \frac{\partial L_{bulk} }{\partial (\partial_i \partial_j g_{ab})}
=[(D/2)-m]L_{bulk}
\end{equation} 
Substituting this in \eq{qF}, transferring these terms to left hand side and using $L\sqrt{-g}-L_{bulk}=L_{sur}$, we get the result in \eq{result1}.

The result in \eq{result1} is the appropriate generalisation of \eq{holorel} in the case of Einstein-Hilbert action and has a similar (generalised) ``$d(qp)$" structure. We shall now turn to the task of connecting up the surface term to horizon entropy so as to provide a thermodynamic interpretation.

\section{Surface term and the entropy of the horizon}
\label{sec:entro}

Surface terms in actions sometimes assume special significance in a theory and this is particularly true for Einstein-Hilbert action. In this case, one can relate the surface term to the entropy of the horizons, if the solution possesses bifurcation horizon. This is well-known in the case of black hole horizons. More generally, if the metric near the horizon can be approximated as a Rindler metric, then one can obtain the general result that the entropy per unit transverse area is $1/4$.
To see this, we only need to evaluate the surface contribution
\begin{equation}
S_{sur} = 2 \int d^D x \, \partial_c \left[ \sqrt{-g} Q^{abcd}\Gamma_{abd} \right]=2 \int d^D x \, \partial_c \left[ \sqrt{-g} Q^{abcd} \partial_b g_{ad}\right]
\label{ehsuren}
\end{equation}
for a metric in the Rindler approximation:
\begin{equation}
ds^2 = - \kappa^2 x^2 dt^2 + dx^2 + dx_\perp^2
\end{equation} 
where $x^m_\perp$ demotes $(D-2)$ transverse coordinates. For the static metric, the time integration in \eq{ehsuren} is trivial and involves multiplication by the range of integration.
Since the surface gravity of the horizon (located at $x=0$) is $\kappa$, the the natural range for time integration is
 $(0,\beta)$ where $\beta=2\pi/\kappa$. (This is most easily seen in the Euclidean sector in which there is a natural periodicity.) Further, it is easy to verify that only the $Q_{0x}^{\phantom{0x}0x}$ term contributes. Then, a
simple calculation shows that
\begin{equation}
S_{sur} = 8\pi \int_H d^{D-2}x_\perp \left( Q_{0x}^{\phantom{0x}0x}\right)
\label{res1}
\end{equation} 
In evaluating this contribution, the $x$ integral in \eq{ehsuren} will range from some $x=a$
to $x=b$ and the result will depend on the behaviour of the integrand at both limits. 
What we have evaluated in \eq{res1} is contribution of the integral from \textit{one} surface, which is taken to be the location of the horizon.
Our Rindler approximation is valid only near the horizon and one cannot say anything about the other contribution without knowing the detailed form of the metric. For example, if the second limit is at infinity, one needs to know whether the metric is asymptotically flat etc. We need not bother about these issues by evaluating the result on the horizon alone, indicated by the subscript $H$ in \eq{res1}.
In the case of Einstein-Hilbert action
$
Q^{abcd} = (1/32\pi) \left[g^{ac}g^{bd} - g^{ad}g^{bc}\right]
$
so that
\begin{equation} 
Q_{0x}^{\phantom{0x}0x} = -\frac{1}{32\pi}; \qquad
S_{\rm sur} = - \frac{1}{4}\, \mathcal{A}_\perp
\end{equation}
as expected. (The minus sign arises because of the Minkowski signature we are working with).
We will now show that this result continues to hold for  Lanczos-Lovelock lagrangians with our definition of $L_{sur}$ thereby providing a  thermodynamic underpinning for our holographic separation of  Lanczos-Lovelock lagrangians.

In the next subsection, we shall provide a proof by comparing contribution of the surface term on the horizon with the Noether charge for these spacetimes. In Sec.~\ref{sec:horizon} we will give a more direct and explicit calculation in the case static, spherically symmetric, solution. 

\subsection{Surface term and the Noether charge}

To do this we need an expression for the entropy of the horizon in a general context when the lagrangian depends of $R^a_{\phantom{a}bcd}$ in a non-trivial manner. Such a formula has been provided by Wald in ref. \cite{noether} and can be expressed as a integral over $P^{abcd}$ on the horizon, evaluated on-shell. It can also been shown \cite{noether} that this definition is equivalent to
interpreting  entropy as the Noether charge associated with diffeomorphism invariance. We shall briefly summarize this approach and use this definition.

To define the Noether charge associated with the diffeomorphism invariance, let us consider the variation $x^a\to x^a + \xi^a$ under which the metric changes by 
$\delta g_{ab} = - (\nabla_a \xi_b + \nabla_b \xi_a)$. The change in the action, when
evaluated on-shell, is contributed only by the surface term so that we have the 
relation
\begin{equation}
\delta_\xi A\big|_{\rm on\ shell} = - \mes{D}{} \nabla_a (L\xi^a) 
                                  = \mes{D}{} \nabla_a(\delta_\xi V^a)
\end{equation} 
[The subscript $\xi$ on $\delta_\xi....$ is a reminder that we are considering the changes due to a particular kind of variation, viz. when the metric changes by $\delta g_{ab} = - (\nabla_a \xi_b + \nabla_b \xi_a)$.]
This leads to the conservation law $\nabla_a J^a =0$ with $J^a = L \xi^a + (\delta_\xi V^a)
\equiv \nabla_b J^{ab}$ with the last equality defining the antisymmetric tensor
$J^{ab}$.
For a lagrangian of the type $L=L(g^{ab},R^a_{bcd})$ direct computation using \eq{deltaa} shows that $J^{ab}$ is given by (also see \cite{defj})
\begin{equation}
J^{ab} = - 2 P^{abcd} \nabla_c \xi_d + 4 \xi_d \left(\nabla_c P^{abcd}\right)
\label{noedef}
\end{equation} 
with $P_{abcd}\equiv (\partial L/\partial R^{abcd})$.
We shall  confine ourselves to  Lanczos-Lovelock type lagrangians for which 
 \begin{equation}
L = \frac{1}{m} R^{abcd} \left( \frac{\partial L}{\partial R^{abcd}}\right) \equiv R^{abcd} Q_{abcd}
\end{equation} 
with $\nabla_a P^{abcd} =\nabla_a Q^{abcd} =0$ so that $J^{ab} = - 2 P^{abcd} \nabla_c \xi_d$.

We want to evaluate the Noether charge corresponding to the current $J^a$ for a static metric with a bifurcation horizon and a killing vector field $\xi^a = (1, \textbf{0})$. The location of the horizon is given by the vanishing of the norm $\xi^a\xi_a = g_{00},$ of this  killing vector.
Using these facts as well as the relations  $\nabla_c\xi^d = \Gamma^d_{c0}$ etc., we find that
$J^{ab} = 2P_{d}^{\phantom{d}0ab}\Gamma^d_{c0}$. Therefore the  Noether charge is given by
\begin{equation}
\mathcal{N} = \mes{D-1}{t} J^0 = \mes{D-1}{t} \frac{1}{\sqrt{-g}}\partial_b(\sqrt{-g} J^{b0})
=\mes{D-2}{t,r_H} J^{r0}
\end{equation} 
in which we have ignored the contributions arising from $b=$ transverse directions. This is
justifiable when transverse directions are compact or in the case of Rindler approximation
when nothing changes along the transverse direction. In the radial direction, we have again confined to the contribution at $r=r_H$ which is taken to be the location of the horizon.
Using 
$Q^{r0} = 2 P^{dcr0} \Gamma_{dc0} = -2 P^{dcr0}\partial_d g_{c0}$ we get
\begin{equation}
\mathcal{N} = -2 \mes{D-2}{t,r_H} P^{dcr0}\partial_d g_{c0} = 2m \mes{D-2}{t,r_H} Q^{cdr0}\partial_d g_{c0}
\label{n}
\end{equation} 
Note that the dimension of $\mathcal{N}$ is $L^{D-3}$ which is area of transverse dimensions
divided by a length. Entropy, which has the dimensions of transverse area, is given by the product of $\mathcal{N}$ and the interval in time integration. If the surface gravity of the horizon is $\kappa$, the time integration can be limited to the range $(0,\beta)$ where $\beta=2\pi/\kappa$. The entropy, computed from the Noether charge approach is thus given by
\begin{equation}
S_{Noether}=\beta\mathcal{N} = 2\beta m \mes{D-2}{t,r_H} Q^{cdr0}\partial_d g_{c0}
\label{s}
\end{equation}  

We will now show that this is the same  result one obtains by evaluating our surface term on the horizon except for a proportionality constant. In the stationary case, the contribution of surface term on the horizon is given by
\begin{equation}
S_{sur} = 2 \int d^D x \, \partial_c \left[ \sqrt{-g} Q^{abcd} \partial_b g_{ad}\right]
= 2\int dt\mes{D-2}{r_H} Q^{abrd} \partial_b g_{ad}
\end{equation} 
Once again, taking the integration along $t$ to be in the range $(0,\beta)$ and ignoring
transverse directions, we get
\begin{equation}
S_{sur} = 2\beta \mes{D-2}{r_H} Q^{abr0} \partial_b g_{a0}
\end{equation} 
Comparing with \eq{n}, we find that 
\begin{equation}
S_{Noether} = mS_{sur}
\end{equation} 
The overall proportionality factor has a simple physical meaning. \eq{mlsur} tells us that the quantity $mL_{sur}$, rather than $L_{sur}$, which has the $``d(qp)"$ structure and it is this particular combination which plays the role of entropy, as to be expected \cite{teitel}.

\subsection{Direct calculation of horizon entropy from surface term}\label{sec:horizon}

In this section, we shall provide a brief outline of an explicit computation of the contribution of the surface term on a horizon and show that it is equal to the standard expression for entropy, computed previously in the literature for  Lanczos-Lovelock lagrangians. To this end, we
 will consider a metric in D-dimensional spacetime in, static, isotropic coordinates with the form
\begin{equation}
ds^{2}= -b(r) dt^{2} + b^{-1}(r) dr^{2} + r^{2}g_{mn}(x) dx^{m}dx^{n}
\end{equation} 
In general, a static, spherically symmetric metric can have different functions describing $g_{00}$ and $g_{rr}$. For our purpose we have assumed $g_{00}g_{rr}=-1$ since many solutions relevant to us fall in this category and it simplifies the calculations.

We will now  evaluate the surface term for the off-shell metric discussed above in the Euclidean spacetime. Let the integrand (the $L_{sur})$ of the surface term be $\partial_{c}P^{c}$. On integration over the radial direction,  this will have two contributions: one from the horizon, $P^r(r_+)$ where the horizon is at $r=r_+$ and the other from the surface at infinity $P^r(\infty)$.  We will again concentrate on the contribution from the horizon.
Let $\Sigma$ be the surface $r=r_{+}+\epsilon$. Then we need to compute:
\begin{equation} 
I_{\epsilon}\equiv\int d\Sigma\,  n_{r}P^{r}=\sqrt{b^{'}(r_{+})\epsilon}\int_{0}^\beta dt\, r_{+}^{D-2}\int d^{D-2} \Omega\, n_{r}P^{r};\quad
\beta= \frac{ 4\pi}{ b^{'}(r_{+})} 
\end{equation}
We have used the measure $d\Sigma$ appropriate for our metric, restricted the  range of integration of $t$ to $(0,\beta)$ as explained earlier and   used the fact that the normal  to $\Sigma$ has the nonvanishing component
  $n_{r}=-1/\sqrt{b^{'}(r_{+})\epsilon}$.
The horizon contribution arises from the limit of $\epsilon \to 0$. The $\sqrt{\epsilon}$ term in the measure cancels with $\sqrt{\epsilon}$ term in the normal.
Further, it can be verified that  $P^{r}$ is regular at the horizon. So the contribution to the surface term from the horizon is 
\begin{equation}
I_{+}=\frac{ 4\pi}{ b^{'}(r_{+})} r_{+}^{D-2}P^{r}(r_{+}) \Sigma_{D-2}
\end{equation} 
 where $\Sigma_{D-2}$ is the (dimensionless) volume of $S^{D-2}$. 
Evaluating  $P^{r}(r)= 2Q^{abrd}\partial_{b}g_{ad}$ explicitly for our metric, we find that 
\begin{equation}
I_{+}=8 \frac{ \pi}{ b^{'}(r_{+})} \Sigma_{D-2} \left[Q^{rtrt}\partial_r g_{tt} + Q^{rmrm}\partial_{r}(r^{2}g_{mm}\right]
\label{iplus}
\end{equation}
This result is general in the sense that we have not assumed  anything about $Q^{abcd}$ so far.

We will now specialize to the
  Lanczos-Lovelock type lagrangian, for which explicit evaluation shows that non-zero components near horizon  gives:
\begin{eqnarray}
Q^{trrt}=\frac{1}{16\pi }\frac{\space ^{D-2}L_{m-1}}{2 r^{2(m-1)}} + \mathcal{O}(b);\quad
Q^{mrmr}=\mathcal{O}(b)\quad
Q^{tmtm}=\mathcal{O}(b)
\label{theqs}
\end{eqnarray} 
where $\space ^{d-2}L_{m-1}$ is the  Lanczos-Lovelock lagrangian of degree $(m-1)$ evaluated on the horizon.
(The vanishing of some of the components can be argued from symmetry considerations. Terms with odd number of $t$'s will vanish in a static situation,  because of time reversal symmetry. Similarly rotational invariance will forbid terms which have odd number of  transverse coordinates. Rotational invariance also implies that $Q^{rmrn}=0$ if $m\neq n$.). 
Using \eq{theqs} in \eq{iplus} we get the final result to be:
\begin{equation}
I_{+}= -\frac{r_{+}^{D-2m}}{4 }\Sigma_{D-2}\; (\space^{D-2}L_{m-1})
\end{equation}
This is just $(1/m)$ times the Wald's entropy (with a minus sign due to the choice of Minkowski signature) and has been computed in the literature before (see e.g.\cite{wen}). In fact whenever $Q^{rmrm}$ vanishes at the horizon, the contribution of the horizon to the surface term is  $(1/m)$ times the Wald's entropy.

Finally we would like to make a comment on the general covariance of the result.  It is easy to show that, if one changes coordinates from $x^a$ to $y^a$ the results will differ by a term that is proportional to:
\begin{equation}
\sqrt{b^{'}(r_{+})\epsilon}\int_{0}^{ \beta }  dt r^{D-2}\int d^{D-2} \Omega\ n_{r}\left(2Q^{abrd}(x)g_{ea}(x)\left(\frac{\partial^2 y^{e}}{\partial x^{c} \partial x^{d} }\right)\left(\frac{\partial x^{c}}{\partial y^b}\right)\right)
\end{equation} 
This term has to evaluated at the  horizon as far a the entropy computation is concerned. On Euclidean continuation, the horizon maps to the origin. For the subset of coordinate transformations (a) which are  regular at the origin and  (b) in which  the transformed coordinates also are like polar coordinates near the origin,  this extra contribution vanishes at the horizon. This is because   $\partial^2 y^{e}/\partial x^{c} \partial x^{d} $  vanishes at the origin, since the only allowed transformation at the origin is a spacetime independent scaling of $r$ and $t$.

\section{Conclusions}
\label{sec:conclu}

Our key conclusions are summarized in the Table, listed  from the most general results to the special case as we proceed down. The title line defines the lagrangian we consider which, under the condition in (1), is generally covariant and has a specific separation in to surface and bulk terms. The most general results are in the first row, which does not assume any structure about $Q_a^{\phantom{a}bcd}$ other than that 
$\nabla_cQ_a^{\phantom{a}bcd}=0$. These relations in the Table show that one can determine $L_{sur}$ and $L_{bulk}$ in terms of each other provided we treat $\Gamma^a_{bc}$ as independent during the differentiation etc., as explained in Sec. \ref{sec:ehaction}. The next row deals with lagrangians which are of  Lanczos-Lovelock type [which satisfy \textit{both} conditions (1) and (2)]. In addition to the results in the previous row, we obtain two more results expressing $L_{sur}$ in terms of $L$ or $L_{bulk}$. The ``$d(qp)$ structure" is obvious in this case. The last row
discusses the well known Einstein-Hilbert lagrangian which has been our reference point.
In this case, the results for  Lanczos-Lovelock lagrangian with ($m=1$), of course, continues to
be valid; but, in addition, we can simplify one of the relations further.

As we discussed before, the surface term (even in the most general case) has a ``$d(qp)$ structure". In the lagrangian picture we have adopted throughout the paper, we treat
all the $g_{ab}$s at the same footing. However, we know that in any generally covariant 
theory the choice of coordinates puts $D$ conditions on the $g_{ab}$ which could be 
conveniently taken to be on $g_{00}$ and $g_{0\alpha}$. Though the Hamiltonian structure
for an arbitrary generally covariant lagrangian is complicated (and --- as far as we 
know --- not fully worked out at the same level as, say, ADM description in general relativity),
the contribution of the surface term on $t=$ constant surfaces will only depend on 
$g_{ab} [\partial L /\partial (\partial_0 g_{ab})]$. If one can impose a gauge condition
that $g_{00} =1$ and $g_{0\alpha} =0$, then this will give the standard canonical momenta
corresponding to the dynamical variables $g_{\alpha\beta}$ in the Hamiltonian language.
This is however a rather formal statement in the absence of a fully developed Hamiltonian
formulation for the  Lanczos-Lovelock lagrangian.

\begin{center}
\begin{tabular}{ll}
\toprule
\noalign{\smallskip}
\multicolumn{2}{c}{Holographic relations in Lagrangians}    \\
\noalign{\bigskip}
\multicolumn{2}{c}{$\sqrt{-g}L=\sqrt{-g}Q_a^{\phantom{a}bcd}R^a_{\phantom{a}bcd} =2\partial_c\left[\sqrt{-g}Q_a^{\phantom{a}bcd}\Gamma^a_{bd}\right]
+2\sqrt{-g}Q_a^{\phantom{a}bcd}\Gamma^a_{dk}\Gamma^k_{bc}\equiv L_{\rm sur} + L_{\rm bulk} $}  \\ 
\noalign{\smallskip}
\hline \hline 
\noalign{\smallskip}
(1) \quad$\nabla_cQ_a^{\phantom{a}bcd}=0$ & 
\begin{minipage}[l]{4in}
\begin{eqnarray*}
&&L_{\rm sur} = -\partial_p \left(\delta^q_r \frac{\partial L_{\rm bulk}}{\partial \Gamma^q_{pr}}
\right)\\
&&L=\frac{1}{2}R^a_{\phantom{a}bcd}\left(\frac{\partial V^c}{\partial \Gamma^a_{bd}}\right);
\quad
L_{bulk}=\sqrt{-g}\left(\frac{\partial V^c}{\partial \Gamma^a_{bd}}\right)
\Gamma^a_{dk}\Gamma^k_{bc}
\end{eqnarray*}
\end{minipage}\\
\noalign{\medskip}
\hline\noalign{\smallskip}
 (2) \quad ${\displaystyle{Q_a^{\phantom{a}bcd} = \frac{1}{m} \frac{\partial L }{\partial R^a_{\phantom{a}bcd}} }}$&
 \begin{minipage}[l]{4in}
\begin{eqnarray*}
 &&L_{\rm sur} = -\partial_p \left(\delta^q_r \frac{\partial L_{\rm bulk}}{\partial \Gamma^q_{pr}}
\right) \\
&&L=\frac{1}{2}R^a_{\phantom{a}bcd}\left(\frac{\partial V^c}{\partial \Gamma^a_{bd}}\right);
\quad
L_{bulk}=\sqrt{-g}\left(\frac{\partial V^c}{\partial \Gamma^a_{bd}}\right)
\Gamma^a_{dk}\Gamma^k_{bc}\\
&&[(D/2) - m]L_{sur} =-\partial_i \left[ g_{ab} \frac{\delta L_{bulk}}{\delta (\partial_i g_{ab})}
  +\partial_jg_{ab} \frac{\partial L_{bulk}}{\partial (\partial_i \partial_jg_{ab})}
  \right] \\
  && mL_{sur} = -\partial_i\left[g_{ab} \frac{\delta L}{\delta (\partial_i g_{ab})}\right]
  \end{eqnarray*}
  \end{minipage}\\
  \noalign{\medskip}
\hline\noalign{\smallskip}
 (3) \quad ${\displaystyle{Q_a^{\phantom{a}bcd}=\frac{1}{2}(\delta^c_ag^{bd}-\delta^d_ag^{bc}) }}$&
  \begin{minipage}[l]{4in}
  \begin{eqnarray*}
  &&L_{\rm sur} = -\partial_p \left(\delta^q_r \frac{\partial L_{\rm bulk}}{\partial \Gamma^q_{pr}}
\right) \\
&&L=\frac{1}{2}R^a_{\phantom{a}bcd}\left(\frac{\partial V^c}{\partial \Gamma^a_{bd}}\right);
\quad
L_{bulk}=\sqrt{-g}\left(\frac{\partial V^c}{\partial \Gamma^a_{bd}}\right)
\Gamma^a_{dk}\Gamma^k_{bc}\\
&&L_{\rm sur}=-\frac{1}{[(D/2)-1]}\partial_i\left(g_{ab}\frac{\partial L_{\rm bulk}}{\partial (\partial_ig_{ab})}\right)\\
&&  L_{sur} = - \partial_i\left[g_{ab} \frac{\delta L}{\delta (\partial_i g_{ab})}\right]
\end{eqnarray*}
\end{minipage}\\
\noalign{\smallskip}
\botrule
\end{tabular}
\end{center}

\section*{Acknowledgment}

The authors thank R. Gopakumar, A. Paranjape and Sudipta Sarkar for discussions 
 and for comments on the manuscript. A. Mukhopadyay thanks IUCAA
for  hospitality during a visit when part of this work was done.

\appendix
\section{Proofs for different relations}\label{sec:mainapp}

In this Appendix we outline the proofs of different equations stated in the text for the sake of completeness.

\subsection{Proof of \eq{gensq} and \eq{simple}}\label{sec:appone}

Consider a lagrangian of the form $L= \sqrt{-g}Q_{a}^{\phantom{a}bcd}R^{a}_{\phantom{a}bcd}$ in which $Q_{a}^{\phantom{a}bcd}$
has the algebraic symmetries of curvature tensor. Expressing $R^{a}_{\phantom{a}bcd}$
in terms of $\Gamma^i_{jk}$ and using the antisymmetry of $Q_{a}^{\phantom{a}bcd}$
in $c$ and $d$, we can write
\begin{eqnarray}
L &=& \sqrt{-g}Q_{a}^{\phantom{a}bcd}R^{a}_{\phantom{a}bcd} 
  = 2 \sqrt{-g}Q_{a}^{\phantom{a}bcd} (\partial_c \Gamma^a_{db} +   \Gamma^a_{ck}\Gamma^k_{db})\nonumber\\
  &=& 2 \sqrt{-g}Q_{a}^{\phantom{a}bcd}\Gamma^a_{ck}\Gamma^k_{db} 
   + 2\partial_c\left[ \sqrt{-g}Q_{a}^{\phantom{a}bcd}\Gamma^a_{db}\right]
   - 2\Gamma^a_{db}\partial_c[ \sqrt{-g}Q_{a}^{\phantom{a}bcd}]\nonumber\\
 &=& 2 \sqrt{-g}Q_{a}^{\phantom{a}bcd}\Gamma^a_{ck}\Gamma^k_{db} 
   + 2\partial_c\left[ \sqrt{-g}Q_{a}^{\phantom{a}bcd}\Gamma^a_{db}\right]
   - 2\sqrt{-g}\Gamma^a_{db}\partial_c Q_{a}^{\phantom{a}bcd} 
   -  2\sqrt{-g}\Gamma^a_{db}\Gamma^j_{cj} Q_{a}^{\phantom{a}bcd} 
   \label{eone}
\end{eqnarray} 
We now express $ \partial_c Q_{a}^{\phantom{a}bcd} $ in terms of 
$\nabla_c Q_{a}^{\phantom{a}bcd} $ to obtain 
\begin{equation}
\Gamma^a_{db}\partial_c Q_{a}^{\phantom{a}bcd} = \Gamma^a_{db} \nabla_c
Q_{a}^{\phantom{a}bcd} - \Gamma^a_{db} \Gamma^b_{kc}Q_{a}^{\phantom{a}kcd}
 + \Gamma^a_{db} \Gamma^k_{ac}Q_{k}^{\phantom{k}bcd}
 - \Gamma^a_{db} \Gamma^c_{kc}Q_{a}^{\phantom{a}bkd}
 \label{etwo}
\end{equation} 
Substituting \eq{etwo} into \eq{eone} we notice that two pairs of the terms cancel out
leaving the result
\begin{equation}
 \sqrt{-g}Q_a^{\phantom{a}bcd}R^a_{\phantom{a}bcd}
 =2\partial_c\left[\sqrt{-g}Q_a^{\phantom{a}bcd}\Gamma^a_{bd}\right]
+2\sqrt{-g}Q_a^{\phantom{a}bcd}\Gamma^a_{dj}\Gamma^j_{bc}-2\sqrt{-g}\Gamma^a_{bd}\nabla_cQ_a^{\phantom{a}bcd}
\end{equation} 
This is essentially our result in \eq{simple}. Since $\nabla_cQ_a^{\phantom{a}bcd}=0$
for the Einstein-Hilbert action, we get \eq{gensq}.

\subsection{Connecting up \eq{holorel} and \eq{holo}}\label{sec:apptwo}

In the text we proved that, for any lagrangian of the form $L= \sqrt{-g}Q_{a}^{\phantom{a}bcd}R^{a}_{\phantom{a}bcd}$ with $\nabla_cQ_{a}^{\phantom{a}bcd}=0$ there is a natural separation of lagrangian into bulk and surface terms, related by:
\begin{equation}
L_{sur} = -\partial_m \left(\delta^p_n\frac{\partial L_{bulk}}{\partial \Gamma^{p}_{mn}}\right)
\end{equation}
The key caveat in this relation is that one needs to treat all components of $\Gamma^{p}_{mn}$ as independent while differentiating and use a particular ordering of indices in the original expression. The purpose of this subsection is to recast this relation in terms of the derivatives of the metric and show the rather special nature of Einstein-Hilbert lagrangian. To convert 
this into a relation involving the partial derivatives of the metric we use the result:
\begin{equation}
\frac{\partial \Gamma^{p}_{mn}}{\partial (\partial_{a}g_{bc})} = \frac{1}{2} (-g^{pa}\delta_{m}^{b}\delta_{n}^{c} + g^{pb}\delta_{m}^{a}\delta_{n}^{c} + g^{pb}\delta_{m}^{c}\delta_{n}^{a})
\end{equation}
from which we have the operator identity:
\begin{equation}
g_{bc}\frac{\partial }{\partial (\partial_{a}g_{bc})} = \frac{1}{2}(-g^{pa}g_{mn} + \delta^p_n\delta^a_m + \delta^p_n\delta^a_m )\frac{\partial}{\partial \Gamma^{p}_{mn}}
\label{asix}
\end{equation}
So that:
\begin{equation}
g_{bc}\frac{\partial L_{bulk}}{\partial (\partial_{a}g_{bc})} = \frac{1}{2}(-g^{pa}g_{mn} + \delta^p_n\delta^a_m + \delta^p_n\delta^a_m )\frac{\partial L_{bulk}}{\partial \Gamma^{p}_{mn}}
\end{equation}
Using \eq{dldG} to determine $(\partial L_{bulk}/\partial \Gamma^{p}_{mn})$ and manipulating the terms we get, after some algebra:
\begin{equation}
g_{bc}\frac{\partial L_{bulk}}{\partial (\partial_{a}g_{bc})} = -\sqrt{-g}(Q_{p}^{\phantom{p}bad}\Gamma^{p}_{bd} + Q_{p}^{\phantom{p}bpd}g^{la}(\Gamma_{lbd} - \Gamma_{bdl})) 
=\sqrt{-g}(\partial_ng_{ad})(Q^{anid}+g^{in}Q_{p}^{\phantom{p}apd}-g^{ia}Q_{p}^{\phantom{p}npd})
\end{equation}
Note that, in the last two terms the indices of $Q^{abcd}$ are contracted among themselves; hence, in the general case, it is not possible to proceed further and relate this result to $L_{sur}$ directly. The Einstein-Hilbert lagrangian is special in the sense that, for
 $Q_{p}^{\phantom{p}bad} = (1/2) (\delta_p^a g^{bd} - \delta_p^d g^{ba})$, we have $Q_{p}^{\phantom{p}bpd}=(1/2)(D-1)g^{bd}$ and the last two terms can be combined with the first to give \eq{holorel}.

\subsection{Proof of \eq{condition}}\label{sec:appadded}

For the m-th  Lanczos-Lovelock lagrangian
$
L=  R_{a_1 b_1}^{\phantom{a_1 b_1}c_1 d_1}.......R_{a_m b_m}^{\phantom{a_m b_m}c_m d_m}\delta_{c_1 d_1.....c_m d_m}^{a_1 b_1.....a_m b_m}
$ we have the result:
\begin{equation}
P^{ab}_{\phantom{ab}cd} = \frac{\partial L}{\partial R_{ab}^{\phantom{ab}cd}} =  m R_{a_2 b_2}^{\phantom{a_1 b_1}c_2 d_2}.......R_{a_m b_m}^{\phantom{a_m b_m}c_m d_m}\delta_{cdc_1 d_1.....c_m d_m}^{aba_2 b_2.....a_m b_m}
\end{equation}
Therefore,
\begin{equation}
\nabla_a P^{ab}_{\phantom{ab}cd} = m \nabla_a R_{a_2 b_2}^{\phantom{a_1 b_1}c_2 d_2}.......R_{a_m b_m}^{\phantom{a_m b_m}c_m d_m}\delta_{cdc_1 d_1.....c_m d_m}^{aba_2 b_2.....a_m b_m}
\end{equation}
We note that the the derivatives of the curvature tensor appearing in the expression are rendered completely antisymmetric in all the lower indices due to the contraction with the alternating tensor. But 
  Bianchi identity states that $\nabla_{[a}R_{a_2 b_2]}^{\phantom{a_2 b_2}c_2 d_2}=0$ and thus we get  $\nabla_a P^{ab}_{\phantom{ab}cd}=0$. 

\subsection{Proof of \eq{deltaa}}\label{sec:appthree}

Consider the variation of the quantity $L\sqrt{-g}$ where $L$ is a generally covariant
scalar made from $g^{ab}$ and $R^a_{\phantom{a}bcd}$. We can express its variation
in the form 
\begin{equation}
\delta \left(L\sqrt{-g}\right) = \left(\frac{\partial L\sqrt{-g}}{\partial g^{ab}}\right)\, \delta g^{ab} + \left(\frac{\partial L\sqrt{-g}}{\partial R^a_{\phantom{a}bcd}}\right)\, \delta R^a_{\phantom{a}bcd}
=\left(\frac{\partial L\sqrt{-g}}{\partial g^{ab}} \right)\,\delta g^{ab} + \sqrt{-g} P_a^{\phantom{a}bcd}\, \delta R^a_{\phantom{a}bcd}
\label{ethree}
\end{equation} 
The term $P_a^{\phantom{a}bcd}\, \delta R^a_{\phantom{a}bcd}$ is generally covariant
and hence can be evaluated in the local inertial frame using 
\begin{equation}
\delta R^a_{\phantom{a}bcd}
= \nabla_c \left(\delta \Gamma^a_{db}\right) - \nabla_d \left(\delta \Gamma^a_{cb}\right) = \frac{1}{2} \nabla_c\left[
g^{ai}\left(-\nabla_i\delta g_{db} + \nabla_d\delta g_{bi} + \nabla_b\delta g_{di} \right)\right]
- \{\textrm{term with } c\leftrightarrow d\}
\end{equation}
When this expression is multiplied by $P_a^{\phantom{a}bcd}$ the middle term $g^{ai}\nabla_d\delta g_{bi}$ does not contribute because of the anti symmetry of 
$P^{ibcd}$ in $i$ and $b$. The other two terms contribute equally and we get
a similar contribution from the term with $c$ and $d$ interchanged.
Hence we get 
\begin{equation}
P_a^{\phantom{a}bcd}\delta R^a_{\phantom{a}bcd} = 2 P^{ibcd} \nabla_c\nabla_d (\delta g_{di})
\label{efive}
\end{equation} 
Manipulating the covariant derivative, this can be re expressed in the form
\begin{equation}
P_a^{\phantom{a}bcd}\delta R^a_{\phantom{a}bcd} = 2 \nabla_c \left[ P^{ibcd} \nabla_b \delta g_{di}\right] - 2 \nabla_b \left[ \delta g_{di} \nabla_c P^{ibcd} \right] + 2 \delta g_{di}
\nabla_b\nabla_c P^{ibcd}
\end{equation} 
Combining this with the first term in \eq{ethree}  and rearranging the expression,
we get
\begin{equation}
\delta L\sqrt{-g} = \left(\frac{\partial L\sqrt{-g}}{\partial g^{ab}} - 2\sqrt{-g} \nabla^m \nabla^n
P_{amnb} \right)\delta g_{ab} + \sqrt{-g}\nabla_j \left[ 2 P^{ibjd} (\nabla_b \delta g_{di} ) - 2\delta g_{di} \nabla_c P^{ijcd}\right]
\end{equation} 
which is the same as \eq{deltaa}.

\subsection{Proof of \eq{eulerder} and \eq{mlsur}}\label{sec:appfour}

To prove \eq{mlsur} we shall prove that
\begin{eqnarray}
g_{np}\frac{\partial L\sqrt{-g}}{\partial (\partial_m g_{np})}=2\sqrt{-g}[P_a^{\phantom{a}bac}\Gamma^m_{bc}-2P^{nbmd}\Gamma_{ndb}]\nonumber\\
g_{np}\partial_s\frac{\partial L\sqrt{-g}}{\partial (\partial_s\partial_m g_{np})}=2\sqrt{-g}[P_a^{\phantom{a}bac}\Gamma^m_{bc}-P^{nbmd}\Gamma_{ndb}]
\label{efour}
\end{eqnarray} 
The Euler derivative on the left hand side of \eq{mlsur} is the difference between the two quantities evaluated above. On subtraction, two terms on the right hand side cancels out and what remains leads to \eq{eulerder} for a \textit{general} lagrangian. For  Lanczos-Lovelock lagrangians we are led to  \eq{mlsur} when we use $P^{abcd}=mQ^{abcd}$. 

Equation (\ref{efour}) can be proved by direct computation but somewhat quicker route is the following: We begin by noting that $(\partial_m g_{np})$ and $(\partial_s\partial_m g_{np})$ occurs in $L$ only through $R^a_{\phantom{a}bcd}$. So if we keep $\delta g_{ab}=0$ but vary $\partial_m g_{np}$ and $\partial_s\partial_m g_{np}$ in $L$ and get an expression of the form:
\begin{equation}
\delta L=A^{mnp}\delta(\partial_m g_{np})+B^{smnp}\delta(\partial_s\partial_m g_{np})
\end{equation} 
we can read of the terms we need in \eq{efour}. To do this we start with \eq{efive} which gives, when $\delta g_{ab}=0$:
\begin{equation}
\delta L=\left(\frac{\partial L}{\partial R^a_{\phantom{a}bcd}}\right)\, \delta R^a_{\phantom{a}bcd}=P_a^{\phantom{a}bcd}\delta R^a_{\phantom{a}bcd} = 2 P^{ibcd} \nabla_c\nabla_d (\delta g_{di})
\end{equation} 
We now expand out $\nabla_c\nabla_d (\delta g_{di})$, using $\delta g_{ab}=0$ repeatedly to get:
\begin{equation}
\delta L=2P^{ibcd}[\delta(\partial_b\partial_c  g_{di}) -\Gamma^k_{db} \delta(\partial_c g_{ki}) 
 -\Gamma^k_{ic} \delta(\partial_b g_{dk}) -\Gamma^k_{bc} \delta(\partial_k g_{di})]
 \label{sevenfive}
\end{equation} 
We have also used the fact that when $\delta g_{ab}=0,\delta \partial_c g_{ab}\neq 0,$ we can write $\nabla_j\delta g_{ab}=\partial_j\delta g_{ab}=\delta\partial_j g_{ab}$ etc. We can now read off $\partial L\sqrt{-g}/\partial (\partial_m g_{np})$ and 
$\partial L\sqrt{-g}/\partial (\partial_s\partial_m g_{np})$ from \eq{sevenfive} since $\sqrt{-g}$ goes for a ride. Contracting with $g_{np}$ and using the symmetries immediately gives the first of the equations in \eq{efour} as well as the result
\begin{equation}
g_{np}\partial_s\frac{\partial L\sqrt{-g}}{\partial (\partial_s\partial_m g_{np})}=2g_{np}\partial_s(\sqrt{-g}P^{psmn})
\label{esix}
\end{equation} 
Finally we use the fact that, when $\nabla_cP^{abcd}=0$, we have the relation:
\begin{equation}
\partial_c[\sqrt{-g}P^{abcd}]=-\sqrt{-g}[\Gamma^a_{kc}P^{kbcd}+\Gamma^b_{kc}P^{akcd}]
\end{equation} 
Using this to simplify the right hand side of \eq{esix} leads to  \eq{efour}.


\begin{thebibliography}{99}
\bibitem{ps1}
One possible way out is to introduce a dynamical variable $v=\dot q$ in a lagrangian $L(\ddot q,\dot q,q)$ to obtain $L(\dot v,v,q)$ which have only upto first derivatives of variables. But now $q$ becomes a contrained variable with vanishing canonical momentum. One can avoid this by eliminating only a few of the $\dot q$ in favour of $v$ and obtain a lagrangian of the form $L(\dot v,v,\dot q,q)$ but clearly there are inherent ambiguities in this. So a general lagrangian $L(\ddot q,\dot q,q)$ leads to certain new difficulties in describing the dynamics and posing the initial value problem.
\end{thebibliography}
\end{document}